\documentclass[12pt]{article}
\usepackage{amssymb}

\def\be{\begin{equation}}
\def\ee{\end{equation}}
\def\bes{\begin{eqnarray}}
\def\ees{\end{eqnarray}}
\def\ba{\begin{array}}
\def\ea{\end{array}}
\def\a{\alpha}

\def\d{\delta}

\def\o{\omega}

\def\t{\theta}
\def\p{\partial}
\def\w{\wedge}
\def\I{\mathcal{I}}

\def\2{{1\over 2}}
\def\4{{1\over 4}}
\def\tr{\mathrm{tr}}

\catcode`\@=11
\def\@citex[#1]#2{%
\if@filesw \immediate \write \@auxout {\string \citation {#2}}\fi
\@tempcntb\m@ne \let\@h@ld\relax \def\@citea{}%
\@cite{%
  \@for \@citeb:=#2\do {%
    \@ifundefined {b@\@citeb}%
      {\@h@ld\@citea\@tempcntb\m@ne{\bf ?}%
      \@warning {Citation `\@citeb ' on page \thepage \space undefined}}%
      {\@tempcnta\@tempcntb \advance\@tempcnta\@ne%
      \@tempcntb\number\csname b@\@citeb \endcsname \relax%
      \ifnum\@tempcnta=\@tempcntb 
        \ifx\@h@ld\relax%
          \edef \@h@ld{\@citea\csname b@\@citeb\endcsname}%
        \else%
          \edef\@h@ld{\ifmmode{-}\else--\fi\csname b@\@citeb\endcsname}%
        \fi%
      \else
        \@h@ld\@citea\csname b@\@citeb \endcsname%
        \let\@h@ld\relax%
      \fi}%
    \def\@citea{,\penalty\@highpenalty\,}%
  }\@h@ld
}{#1}}

\def\@citeb#1#2{{[#1]\if@tempswa , #2\fi}}
\def\@citeu#1#2{{$^{#1}$\if@tempswa , #2\fi }}
\def\@citep#1#2{{#1\if@tempswa , #2\fi}}

\title{Virasoro generators and the dS$_3$/CFT$_2$ correspondence\footnote{Research supported in part by the DoE under grant DE-FG05-91ER40627.}}
\author{Scott Ness\footnote{ness@utk.edu}\ \ and George Siopsis\footnote{siopsis@tennessee.edu}\\
\em Department of Physics
and Astronomy, \\
\em The University of Tennessee, Knoxville, \\
\em TN 37996 - 1200, USA.
}
\date{April 2005}
\begin{document}

\maketitle
\vspace{-3.5in}\hfill UTHET-05-0201\vspace{3.5in}

\abstract{We discuss the quantization of a scalar field in three-dimensional asymptotic de Sitter space. We obtain explicit expressions for the Noether
currents generating the isometry group in terms of the modes of the scalar
field and the Liouville gravitational field.
We extend the $SL(2,\mathbb{C})$ algebra of the Noether charges to a full
Virasoro algebra by introducing non-local conserved charges.
The Virasoro algebra has the expected central charge in the weak coupling limit (large central charge $c=3l/2G$, where $l$ is the dS radius and $G$ is Newton's constant).
We derive the action of the Virasoro charges on states in the boundary CFT
thus elucidating the dS/CFT correspondence.}
\newpage
\section{Introduction}

As recent astronomical data suggest~\cite{perlmutter}, we live in a Universe
of positive cosmological constant.
It is therefore imperative that one understand asymptotic de Sitter spaces.
Even though a string theoretical understanding of de Sitter spaces at a quantum
level is still a challenge, there has been significant progress in our
understanding of the dS/CFT corresponence~\cite{strominger,ds,BMS,CK,KV,Klemm,bibus}.
As was argued by Strominger~\cite{strominger}, states in the bulk de Sitter
space are holographically dual to corresponding states on the CFT at the
asymptotic boundary $\mathcal{I}^-$ (infinite past). Further evidence of this
duality was presented in~\cite{BMS}.
In three dimensions, the asymptotic symmetry group is infinite-dimensional,
matching the conformal group on $\mathcal{I}^-$.
The central charge was computed in parallel with the AdS case~\cite{CHvD,BH}
and found to be
\be\label{eq0} c = \frac{3l}{2G} \ee
where $l$ is the dS radius and $G$ is Newton's constant.
This expression was shown to be in agreement with a CFT calculation on the boundary of dS~\cite{CK,KV,Klemm}.
The CFT is a Euclidean Liouville theory and describes the asymptotic dynamics
of three-dimensional gravity with a positive cosmological constant.
Incorporating matter fields does not alter the central charge, however the
matter fields do contribute by renormalizing the coupling constant.
Understanding the CFT dynamics in the presence of matter fields is a challenge.

Here we aim at further elucidating the dS/CFT correspondence for massive scalar
fields in three dimensions. Extending the results of ref.~\cite{BMS},
we obtain explicit expressions for all Noether charges generating the dS$_3$
isometry group in terms of the modes of the scalar field and the Liouville
gravitational field. These are semi-classical expressions valid in the
weak coupling limit (large central charge). Quantum corrections may then be
calculated using perturbation theory.
We then extend the algebra of isometries to a full Virasoro algebra by
introducing non-local conserved charges for the massive field (their Liouville
gravitational field counterparts have already been discussed in~\cite{CK}).
We discuss the action of the Virasoro generators on the boundary CFT states.

Our discussion is organized as follows. We start in section~\ref{sect1} by sketching the derivation of the
salient features of pure three-dimensional gravity~\cite{Witten,AT} whose asymptotic dynamics
is described by a Euclidean Liouville theory~\cite{CK,KV,Klemm}.
In section~\ref{sect2}, we introduce a massive scalar field. Using static coordinates in dS$_3$, we expand it in modes
in the southern and northern diamonds, respectively.
We also obtain its asymptotic form on the dS boundary, $\mathcal{I}^-$.
In section~\ref{sect3}, we express the Noether charges in terms of the modes of
the massive field.
We find expressions which may be easily extended to yield a full Virasoro
algebra. The new charges are conserved and non-local. We show that their central charge vanishes classically.
We also obtain the action of the Virasoro generators on the boundary fields.
In section~\ref{sect4}, we discuss the quantization of the system. We obtain
semi-classical expressions of the Virasoro operators which form an algebra of
central charge given by~(\ref{eq0}) in the weak coupling limit ($c\gg 1$).
We discuss their action on states built from the Euclidean vacuum which is dual to the $SL(2,\mathbb{C})$ invariant CFT vacuum on the boundary.
Finally, in section~\ref{sect5}, we briefly summarize our conclusions.

\section{Asymptotic de Sitter space}\label{sect1}

In this section we discuss the salient features of pure three-dimensional
gravity with a positive cosmological constant. Our discussion is based mostly on refs.~\cite{CK,KV,Klemm}.
de Sitter space emerges as a classical solution. We shall parametrize the metric
using planar coordinates
\be\label{eqpl} ds_{\mathrm{dS}}^2 = - \frac{d\tau^2}{\tau^2} + \tau^2 dzd\bar z \ee
henceforth setting the dS radius
\be\label{eql} l = 1\ . \ee
Three-dimensional gravity is derivable from the Chern-Simons action~\cite{Witten,AT}
\be\label{eqCSa}
  S=-\frac{i}{16\pi G}\int \mathrm{Tr} \left( A\w dA+{2\over 3}A\w A\w A \right) +\frac{i}{16\pi G}\int \mathrm{Tr} \left( \bar{A}\w d\bar{A}+{2\over3}\bar{A}\w\bar{A}\w\bar{A} \right)
\ee
The metric can be obtained from the $SL(2,\mathbb{C})$ vector potentials
\be
  A_\mu = A_\mu^a \tau_a \ \ , \ \ \bar A_\mu = \bar A_\mu^a \tau_a,
\ee
where $\tau_a$ are the $SL(2,\mathbb{C})$ generators
\be
  \tau_0 = \frac{1}{2} \left( \ba{cc} -i & 0\\ 0& i\ea \right) \ \ , \ \
  \tau_1 = \frac{1}{2} \left( \ba{cc}0 & 1\\ 1& 0\ea \right) \ \ , \ \
  \tau_2 = \frac{1}{2} \left( \ba{cc}0 & -i\\ i& 0\ea \right).
\ee
with the normalization condition
\be
  \tr (\tau_a\tau_b) = \frac{1}{2} \eta_{ab}.
\ee
It is advantageous to switch to the new basis
\be
  \tau_\pm = \frac{1}{2} (\tau_1 \mp i\tau_2) \ \ , \ \
  \tau_+ = \frac{1}{2} \left( \ba{cc} 0 & 0\\ 1& 0\ea \right) \ \ , \ \
  \tau_- = \frac{1}{2} \left( \ba{cc} 0 & 1\\ 0& 0\ea \right).
\ee
We can express the vector potential in the new basis explicitly as
\be
  A_\mu = \frac{1}{2} \left( \ba{cc} -iA_\mu^0 & A_\mu^-\\ A_\mu^+& iA_\mu^0\ea \right),
\ee
and similarly for $\bar A_\mu$.

The metric expressed in terms of the vector potential is given as
\be\label{met}
  g_{\mu\nu} = -\frac{1}{2} \tr (A_\mu -\bar A_\mu)(A_\nu-\bar A_\nu).
\ee
For asymptotic de Sitter space,
we need
\be\label{eqmet}
  g_{z\bar z} = \frac{\tau^2}{2} + o(1) \ \ , \ \
  g_{zz} = o(1) \ \ , \ \ g_{\tau\tau} = -\frac{1}{\tau^2} + o(\tau^{-4})\ \ , \ \
  g_{z\tau} = o(\tau^{-3}).
\ee
We can solve the equations of motion for the vector potential, plug the solutions into (\ref{met}), and get explicit corrections to the asymptotically de Sitter metric.  
The equations of motion from the action~(\ref{eqCSa}) yield
\be
  A_\mu = G^{-1} \p_\mu G \ \ , \ \ \bar A_\mu = \bar G^{-1} \p_\mu \bar G
\ee
where
\be G = g(z) M\ \ ,\ \ \bar G = \bar g(\bar z) M^{-1}\ \ , \ \
M = \left( {\sqrt{\tau} \atop 0}\ {0\atop 1/\sqrt{\tau}} \right). \ee
The only dynamical variable is $\hat g = g^{-1} \bar g$. We obtain the components
\be
  A_\tau = - \bar A_\tau = \frac{1}{2\tau} \left( \ba{cc} 1 & 0 \\ 0 & -1 \ea \right) \ \ , \ \ A_{\bar z} = \bar A_z = 0\ ,
\ee
and
\be\label{gauss}
  A_z = - M^{-1} \p_z \hat g \hat g^{-1} M \ \ , \ \ \bar A_{\bar z}= M \hat g^{-1} \p_{\bar z} \hat g M^{-1}.
\ee
Decomposing $\hat g$ {\em \`a la} Gauss (including an appropriate normalization factor),
\be
  \hat g = e^{-\sqrt{2G} \Omega } \left( \ba{cc} e^{2\sqrt{2G} \Omega} + XY & X \\ Y& 1\ea \right)
  \ \ .
\ee
we obtain
\be
  A_z = M^{-1} \left( \ba{cc} \sqrt{2G}\p_z\Omega + e^{-2\sqrt{2G}\Omega}X\p_z Y & \p_z X - 2\sqrt{2G} X\p_z \Omega -e^{-2\sqrt{2G}\Omega} X^2 \p_z Y\\
e^{-2\sqrt{2G}\Omega} \p_z Y& -\sqrt{2G}\p_z\Omega - e^{-2\sqrt{2G}\Omega}X\p_z Y\ea \right)
M.
\ee
and similarly for $\bar A_{\bar z}$.
All three fields $X$, $Y$ and $\Omega$ are independent of the $\tau$ coordinate.
Demanding asymptotic de Sitter space leads to constraints on $X$ and $Y$ of the form
\[
X = i\sqrt{2G} \p_z\Omega \ \ , \ \ Y = -i\sqrt{2G} \p_{\bar z}\Omega
\]
\be\label{eqXY}
e^{-2\sqrt{2G}\Omega} \p_{\bar z} X = i\ \ ,\ \  e^{-2\sqrt{2G}\Omega} \p_z Y = -i\ .
\ee
Using these constraints we arrive at explicit expressions for the $SL(2,\mathbb{C})$ fields solely
in terms of the field $\Omega$
\be
  A_z = \left( \ba{cc} 0 & \frac{2iG}{\tau}\Theta_{zz} \\ -i\tau & 0\ea \right)\ \ ,\ \ 
  \bar A_{\bar z} = \left( \ba{cc} 0 & i\tau\\ -\frac{2iG}{\tau} \Theta_{\bar z\bar z}& 0\ea \right),
\ee
where off-diagonal elements are proportional to the stress-energy tensor of a linear dilaton theory given by
\be
  \Theta_{zz} = (\p_z\Omega)^2 -\frac{1}{\sqrt{2G}}\p_z^2 \Omega\;,
\ \  \Theta_{\bar z\bar z} = (\p_{\bar z}\Omega)^2 -\frac{1}{\sqrt{2G}}\p_{\bar z}^2 \Omega\;,
\ee
The constraints~(\ref{eqXY}) imply that the field $\Omega$ obeys the Liouville equation
\[ 
  \p_z \p_{\bar z} \Omega = \frac{1}{\sqrt{2G}} e^{2\sqrt{2G}\Omega}.
\]
Noether charges for the linear dilaton theory are given by
\be\label{eqLb}
  L_n = \oint_{\mathcal{C}} \frac{dz}{2\pi i} z^{n+1} \Theta_{zz}\ \ , \ \
\bar L_n = -\oint_{\mathcal{C}} \frac{d\bar z}{2\pi i} \bar z^{n+1} \Theta_{\bar z\bar z}
\ee
These charges form Virasoro algebras with central charge
\be\label{eq0c}
  c = \4+\frac{3}{2G} \approx \frac{3}{2G}\;.
\ee
in agreement with~(\ref{eq0}) for $l=1$ (in the limit $G\ll 1$).
The contour $\mathcal{C}$ may be viewed as residing on the boundary of dS$_3$.
For $n=-1,0,1$, they form a $SL(2,\mathbb{C})$ algebra. In the latter case,
they are related to bulk expressions involving the stress-energy tensor $\mathcal{T}_{\mu\nu}$~\cite{BH} by a conservation law~\cite{BMS}
\be\label{eqLbu} L_n = \int_{\Sigma_{\mathcal{C}}} d^2\Sigma^\mu \mathcal{T}_{\mu\nu} \zeta_n^\nu \ \ \ , \ \ n = -1,0,1\ee
where $\zeta_n^\nu$ is the corresponding Killing vector and $\Sigma_{\mathcal{C}}$ is a two-dimensional surface in dS$_3$ whose boundary is the curve $\mathcal{C}$ (similarly for $\bar L_n$).

Using the explicit form of the vector potential and eq.~(\ref{met}), we arrive at an explicit expression for the metric on an asymptotic de Sitter space,
\be\label{g}
  g_{z\bar z} = \frac{\tau^2}{2} + \frac{2G^2}{\tau^2} \Theta_{zz} \Theta_{\bar z\bar z} \ \ , \ \
  g_{zz} = -2G \Theta_{zz} \ \ , \ \
  g_{\bar z\bar z} = -2G \Theta_{\bar z\bar z} \ \ , \ \
  g_{\tau\tau} = -\frac{1}{\tau^2} \ \ , \ \
  g_{z\tau} = g_{\bar z\tau} = 0 \ ,
\ee
clearly satisfying the requirement~(\ref{eqmet}).

\section{Scalar Field}\label{sect2}
Next, we consider a massive scalar field $\Phi$ of mass $m$ living in three-dimensional de Sitter space (dS$_3$).
It is advantageous to parametrize dS$_3$ using static coordinates,
\be
  ds_{\mathrm{dS}}^2 = -(1-r^2)dt^2+\frac{dr^2}{1-r^2}+r^2d\t^2
\ee
instead of the planar coordinates~(\ref{eqpl}) used in the pure gravitational case.
The two coordinate systems are related by a conformal transformation near the
boundary~$\mathcal{I}^-$ (infinite past, $r, \tau\to\infty$)~\cite{BMS},
\be\label{eqconf} z = e^{-iw} \ \ , \ \ w = \theta+it\ee
The field $\Phi$ obeys the wave equation
\be\label{eqwav}
  \frac{1}{r}\p_r \left( r (1-r^2)\p_r\Phi\right) -\frac{1}{1-r^2}\p_t^2\Phi+\frac{1}{r^2}\p_\t^2\Phi =m^2\Phi \ .
\ee
Following ref.~\cite{BMS}, we shall solve this equation in the northern and southern diamonds corresponding to the range $0\le r<1$ and then analytically
continue the solution beyond these domains
aiming at reaching the boundary $\I^-$ which belongs to the past triangle.
We shall present details for the southern diamond; the discussion in the northern diamond is similar.

A complete set of positive frequency solutions in the southern diamond is provided by the
wavefunctions
\be\label{eqphip}
  \phi_{\o j}^S =A_{\o j} f_{\o j}(r)e^{-i\o t+ij\t}\;,
\ee
where $f_{\o j}$ obeys the radial equation
\be\label{waveeqn}
  (1-r^2) f_{\o j}'' +\left(\frac{1}{r}-3r\right) f_{\o j}' +\left(\frac{\o^2}{1-r^2}-\frac{j^2}{r^2}-m^2\right)f_{\o j}=0.
\ee
and we have included a normalization constant $A_{\o j}$, to be determined.
A solution of~(\ref{waveeqn}) which is regular at $r=0$ is given by
\be
  f_{\o j} = r^{|j|}(1-r^2)^{i\o/2} \; F(a_+,a_-;c;r^2)\ \ , \ \
  a_\pm =\2(i\o+|j|+h_\pm)\ \ , \ \
  c =1+|j| \ ,
\ee
where
\be h_\pm = 1\pm i\mu \ \ , \ \ \mu = \sqrt{m^2-1} \ \ , \ \ m^2 > 1 \ . \ee
The choice of normalization constant
\be\label{nc}
  A_{\o j}= e^{i\pi |j|/2}\, \frac{\Gamma(\2(i\o+|j|+h_+))\Gamma(\2(-i\o+|j|+h_+))}{\Gamma(1+|j|)}
\ee
yields an orthonormal set of wavefunctions,
\be\label{phinorm}
  \langle\phi_{\o j}^S|\phi_{\o' j'}^S\rangle = \frac{1}{2\sinh\pi\o} \d(\o-\o')\d_{j,j'} \ ,
\ee
under the inner product
\be\label{innerproduct}
  \langle\phi | \chi\rangle = \2\int_{\Sigma^S}\frac{r drd\t}{1-r^2} \;\phi^*\stackrel{\leftrightarrow}{\p_t}\chi,
\ee
where we are integrating over a spacelike hypersurface $\Sigma^S$ ($t=$~const.) in the southern diamond ($0\le r <1$, $0\le\theta < 2\pi$).

It is useful to find expressions of the negative frequency eigenfunctions in terms
of the positive frequency ones~(\ref{eqphip}).
Using the transformation properties of hypergeometric functions, one can show that
\be\label{phit}
  \phi^{S*}_{\o j} = e^{2i\vartheta_{\o j}} \phi^S_{-\o-j}\ \ , \ \
e^{2i\vartheta_{\o j}} = (-)^{|j|}\; \frac{\Gamma(\2(i\o+|j|+h_-))\Gamma(\2(-i\o+|j|+h_-))}{\Gamma(\2(i\o+|j|+h_+))\Gamma(\2(-i\o+|j|+h_+))}\, .
\ee
This is a phase factor ($\vartheta_{\o j}\in\mathbb{R}$) for frequencies $\omega$ along
the real and imaginary axes.
Notice also that $\vartheta_{-\o\; j} = \vartheta_{\o j} = \vartheta_{\o \; -j}$.
It is of interest to note that $\vartheta_{\o j}$ may also be expressed in terms of $j$ as
\be\label{th-+}
e^{2i\vartheta_{\o j}} = (-)^j\; \frac{\Gamma(\2(i\o+j+h_-))\Gamma(\2(-i\o+j+h_-))}{\Gamma(\2(i\o+j+h_+))\Gamma(\2(-i\o+j+h_+))}\, .
\ee
This is obvious for $j>0$. For $j<0$, it follows from standard Gamma function identities.

A general solution of the wave equation~(\ref{eqwav}) in the southern diamond may be expanded as
\be
  \Phi^S(t,r,\t)=\sum_{j=-\infty}^\infty \int_0^\infty d\o \left( \alpha_{\o j}^S\phi_{\o j}^S + \alpha^{S\dagger}_{\o j}\phi^{S*}_{\o j} \right)\;,
\ee
This may also be written as a single integral over the entire real $\o$-axis as
\be\label{eqPha}
  \Phi^S(t,r,\t)=\sum_{j=-\infty}^\infty \int_{-\infty}^\infty d\o \; \alpha_{\o j}^S\phi_{\o j}^S \;,
\ee
where the negative frequency modes are given by
\be\label{a-+}
  \alpha^S_{-\o \; -j}= e^{2i\vartheta_{\o j}} \alpha^{S\dagger}_{\o j}\ee
on account of~(\ref{phit}).
These modes are related to the annihilation operators in~\cite{BMS} by
\be \alpha_{\o j}^S = \sqrt{\sinh\pi\o}\; a_{\o j}^S \ \ , \ \ \o > 0\ee
due to our choice of normalization condition~(\ref{phinorm}).
They obey the commutation relations
\be\label{eqaad} [ \alpha_{\o j}^S , \alpha_{\o' j'}^S ] = \sinh\pi\o\; e^{2i\vartheta_{\o j}}\; \delta (\o + \o') \delta_{j+j',0}\ee
Strictly speaking, since we are discussing the classical theory, the above should be interpreted in terms of
Poisson brackets,
\be\label{eqconf0} [ \mathcal{A} , \mathcal{B} ] \equiv i\{ \mathcal{A} , \mathcal{B} \}_{\mathrm{P.B.}}\ee
Eq.~(\ref{eqaad}) also holds quantum mechanically, after the modes $\alpha_{\o j}^S$ are promoted to operators.
At slight risk of confusion, we shall use the notation~(\ref{eqconf0}) invariably, pointing out the potential changes under quantization, as needed.

We may use Kruskal coordinates to go beyond the horizon ($r=1$) and into the
past triangle whose boundary is $\mathcal{I}^-$ ($r\to\infty$).
The eigenfunctions in the southern diamond may then be written as linear
combinations of eigenfunctions in the past triangle~\cite{BMS}.
After some algebra, one obtains for the general wavefunction~(\ref{eqPha})
\be
  \Phi^S(r,t,\theta) = \Phi^{S+}(r,t,\theta)+\Phi^{S-}(r,t,\theta)
\ee
where
\bes
  \Phi^{S-}(r,t,\theta) &=&\sum_{j=-\infty}^\infty\int_{-\infty}^\infty \frac{d\o}{\sinh\pi\o} \; \alpha^S_{\o j}\phi_{\o j}^- \;,\nonumber\\
\Phi^{S+} (t,\theta) &=& \Phi^{S-\dagger} (t,\theta) \;,
\ees
and
\be
  \phi_{\o j}^- =B_{\o j} \; r^{-h_-}
\left( 1-\frac{1}{r^2} \right)^{i\o/2}
F(b_{\o j}, b_{\o\; -j};h_-;\frac{1}{r^2}) e^{-i\o t+ij\theta},
\ee
\be
B_{\o j} =  i \Gamma(i\mu) \sin \frac{\pi}{2} (-i\o + j + h_-) \ \ , \ \
b_{\o j} = \frac{1}{2} (i\o + j + h_-) \; . \ee
In deriving the above, we used $i^{|j|} \sin \left( \frac{\pi}{2} |j| + \lambda\right) = \sin \left( \frac{\pi}{2} j+ \lambda \right)$
in order to express the wavefunction entirely in terms of $j$ rather than its absolute value, $|j|$.

Near the boundary $\mathcal{I}^-$, we have
\be \Phi^{S\pm} (r,t,\theta) \sim \frac{\Psi^{S+} (t,\theta)}{r^{h_+}} + \frac{\Psi^{S-} (t,\theta)}{r^{h_-}}\, , \ee
where the boundary wavefunctions are
\bes\label{eqwfb}
\Psi^{S-} (t,\theta) &=& \sum_{j=-\infty}^\infty\int_{-\infty}^\infty \frac{d\o}{\sinh\pi\o}\; B_{\o j}\; \alpha_{\o j}^S \; e^{-i\o t+ij\theta} \;,\nonumber\\
\Psi^{S+} (t,\theta) &=& \Psi^{S-\dagger} (t,\theta) \;.
\ees
Similar expressions hold for boundary wavefunctions $\Psi^{N\pm} (t,\theta)$
derived from the northern diamond.
 
\section{Symmetries and Corresponding Charges}\label{sect3}

The group of isometries of de Sitter space in three dimensions
is generated by the Killing vectors
\bes\label{zeta}
  \zeta_0 &=& \2 (\p_t+i\p_\theta),\nonumber\\
  \zeta_{\pm 1} &=& \2 e^{\pm(t-i\theta)}\left[\frac{\mp r}{\sqrt{1-r^2}}\p_t-\sqrt{1-r^2}\left(\p_r\mp\frac{i}{r}\p_\theta\right)\right],\nonumber\\
  \bar\zeta_0 &=& \2 (\p_t-i\p_\theta)\nonumber\\
  \bar\zeta_{\pm 1} &=& \2 e^{\pm(t+i\theta)}\left[\frac{\mp r}{\sqrt{1-r^2}}\p_t-\sqrt{1-r^2}\left(\p_r\pm{i\over r}\p_\theta\right)\right]\; ,
\ees
forming a $SL(2,\mathbb{C})$ algebra,
\be
  [\zeta_n,\zeta_m]=(n-m)\zeta_{n+m}\ ,\ \ \  [\bar\zeta_n,\bar\zeta_m]=(n-m)\bar\zeta_{n+m}\ ,\ \ \ n,m=-1,0,1
\ee
We can easily construct the corresponding Noether currents and charges,
\be\label{charge}
 Q_n = \int_\Sigma d^2\Sigma\; j^0_n\ ,\ \ \  j^\mu_n = g^{\mu\nu}T_{\nu\a}\zeta^\a_n\;,
\ee
where
\be\label{T}
  T_{\mu\nu}=\p_\mu\Phi\p_\nu\Phi- g_{\mu\nu}\mathcal{L}\ , \ \
\mathcal{L} = \2 g^{\rho\sigma} \p_\rho\Phi \p_\sigma\Phi - \2 m^2 \Phi^2
\ee
and similarly for $\bar Q_n$.
Henceforth, we shall concentrate on the southern diamond and find the corresponding charge $Q_n^S$
by choosing $\Sigma = \Sigma^S$, as in eq.~(\ref{innerproduct}).
The analysis of $Q_n^N$ on the northern diamond is similar. The total charge is
\be\label{eqQtot} Q_n = Q_n^S + Q_n^N\ee
If we integrate by parts and use the wave equation~(\ref{eqwav}), we can massage (\ref{charge}) into the form
\be\label{charge2}
  Q_n^S =\langle\Phi^S|\zeta_n\Phi^S\rangle \, ,
\ee
where the inner product is defined in (\ref{innerproduct}).
The Killing vectors $\zeta_n$ act on the eigenstates $\phi_{\o j}^S$ (eq.~(\ref{eqphip})) in a
particularly simple manner,
\be\label{eq45}
  \zeta_n\phi_{\o j}^S = \frac{1}{2} (i\o+j- nh_-)\phi_{\o+in,j-n}^S\ ,\ \ n=-1,0,1\;.
\ee
This is partly due to our judicious choice of normalization constant~(\ref{nc}) and is shown using standard hypergeometric function identities. Remarkably,
eq.~(\ref{eq45}) is independent of the sign of $j$, even though its derivation
follows different paths in the two cases, $j>0$ and $j<0$, respectively.
We may use it, together with the orthogonality conditions~(\ref{phinorm}),
to express the Noether charges (\ref{charge2}) in terms of creation and annihilation operators.
To arrive at such an expression for $Q_n^S$, let us first write it in the form
\be Q_n^S = \2 \sum_{j=-\infty}^\infty\int_{-\infty}^\infty d\o\; \alpha_{\o j}^S\; \langle\Phi^S|\zeta_n\phi_{\o j}^S\rangle \ee
Using~(\ref{eq45}), this can be written as
\be Q_n^S = \frac{1}{2} \sum_{j=-\infty}^\infty\int_{-\infty}^\infty d\o\; (i\o +j -nh_-)\alpha_{\o j}^S\; \langle\Phi^S|\phi_{\o+ni\; j-n}^S\rangle \ee
Stifting $\o \to \o -ni$, $j\to j+n$, yields
\be Q_n^S = \frac{1}{2} \sum_{j=-\infty}^\infty\int_{-\infty}^\infty d\o\; (i\o +j +nh_+)\alpha_{\o-ni\; j+n}^S\; \langle\Phi^S|\phi_{\o j}^S\rangle \ee
The inner products may be easily evaluated by expanding $\Phi$ in modes (eq.~(\ref{eqPha})).
Using the orthogonality relation~(\ref{phinorm}), we obtain
\be Q_n^S = \frac{1}{4} \sum_{j=-\infty}^\infty\int_{-\infty}^\infty \frac{d\o}{\sinh\pi\o}\; (i\o +j +nh_+)\alpha_{\o-ni\; j+n}^S\; \alpha_{\o j}^{S\dagger} \ee
Shifting back $\o \to \o +ni$, $j\to j-n$, we obtain the equivalent expression
\be\label{eqQ} Q_n^S = \frac{1}{4} \sum_{j=-\infty}^\infty\int_{-\infty}^\infty \frac{d\o}{\sinh\pi\o}\; (i\o +j -nh_-)\alpha_{\o j}^S\; \alpha_{\o+ni\; j-n}^{S\dagger} \ee
The above expressions are deceptively simple; their complexities are revealed in eqs.~(\ref{a-+}) and (\ref{th-+}).
Remarkably, even though they have been derived for $n=-1,0,1$ as generators of isometries, they are well-defined for all $n\in\mathbb{Z}$.
One may be tempted to use this fact to extend the $SL(2,\mathbb{C})$ algebra of isometries to an infinite-dimensional algebra of the conserved charges $Q_n^S$ ($n\in\mathbb{Z}$).
Unfortunately, these charges do not form a closed algebra.

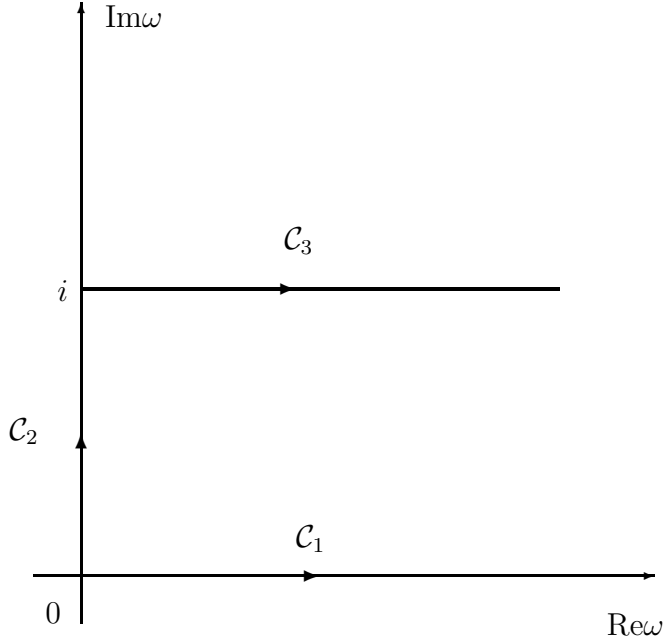
\begin{figure}
\begin{center}
\setlength{\unitlength}{3947sp}%
\begingroup\makeatletter\ifx\SetFigFont\undefined%
\gdef\SetFigFont#1#2#3#4#5{%
  \reset@font\fontsize{#1}{#2pt}%
  \fontfamily{#3}\fontseries{#4}\fontshape{#5}%
  \selectfont}%
\fi\endgroup%
\begin{picture}(4062,4041)(751,-4390)
\thinlines
{\put(1201,-4261){\vector( 0, 1){3900}}
}%
{\put(901,-3961){\vector( 1, 0){3900}}
}%
\thicklines
{\put(1201,-3961){\vector( 1, 0){1500}}
}%
{\put(1201,-2161){\vector( 1, 0){1350}}
}%
{\put(1201,-3961){\vector( 0, 1){900}}
}%
{\put(1201,-3211){\line( 0, 1){1050}}
}%
{\put(2401,-2161){\line( 1, 0){1800}}
}%
{\put(2476,-3961){\line( 1, 0){1725}}
}%
\put(2476,-1936){\makebox(0,0)[lb]{$\mathcal{C}_3$%
}}
\put(2551,-3811){\makebox(0,0)[lb]{$\mathcal{C}_1$%
}}
\put(751,-3136){\makebox(0,0)[lb]{$\mathcal{C}_2$%
}}
\put(1051,-2236){\makebox(0,0)[lb]{$i$%
}}
\put(976,-4261){\makebox(0,0)[lb]{$0$%
}}
\put(4501,-4336){\makebox(0,0)[lb]{Re$\omega$%
}}
\put(1351,-511){\makebox(0,0)[lb]{Im$\omega$%
}}
\end{picture}
\end{center}
\caption{\label{fig1} Contour deformation for $Q_n^{S-}$ (eq.~(\ref{eqQ-}))
for $n=1$.
$\delta Q_n^S$ (eq.~(\ref{eqdQ})) corresponds to an integral
over the segment $\mathcal{C}_2$.}
\end{figure}

To remedy this, let us split the integral~(\ref{eqQ}) into positive and negative frequencies,
\be Q_n^S = Q_n^{S+} + Q_n^{S-} \ee
Using eq.~(\ref{a-+}), we may write $Q_n^{S-}$ in terms of positive frequencies as
\be Q_n^{S-} = \frac{1}{4} \sum_{j=-\infty}^\infty\int_0^\infty \frac{d\o}{\sinh\pi\o}\; (i\o +j +nh_-) e^{2i(\vartheta_{\o j}- \vartheta_{\o-ni\; j+n})} \alpha_{\o j}^{S\dagger}\; \alpha_{\o-ni\; j+n}^S\ee
It is straightforward to deduce from the definition~(\ref{th-+}),
\be\label{eq64} (i\o +j +nh_-) e^{2i(\vartheta_{\o j}- \vartheta_{\o-ni\; j+n})} = (-)^n(i\o +j+nh_+) \ \ \ , \ \ n=-1,0,1\ee
It is worth mentioning that eq.~(\ref{eq64}) does not hold for any other $n\in\mathbb{Z}$.
It follows that
\be\label{eqQ-} Q_n^{S-} = \frac{1}{4} \sum_{j=-\infty}^\infty\int_0^\infty \frac{d\o}{\sinh\pi\o}\; (i\o +j +nh_+)\; \alpha_{\o j}^{S\dagger}\; \alpha_{\o-ni\; j+n}^S\ee
for $n=-1,0,1$.
Shifting $\o\to\o + ni$, $j\to j-n$ yields an expression which matches the one
for $Q_n^{S+}$, as is evident from eq.~(\ref{eqQ}).
However, we need to exercise care in this shift, since the integral is over the
positive real $\omega$-axis.
The shift amounts to a contour deformation in the complex $\o$-plane and yields
an additional contribution (see fig.~\ref{fig1} for $n=1$; $n=-1$ is similar;
for $n=0$ we have $\delta Q_0^S = 0$, trivially),
\be\label{eqdQ} \delta Q_n^S = \frac{1}{4} \sum_{j=-\infty}^\infty\int_0^n \frac{d\lambda}{\sin\pi\lambda}\; (-\lambda +j +nh_+)\; \alpha_{i\lambda j}^{S\dagger}\; \alpha_{i(\lambda-n)\; j+n}^S\ee
This can be seen to vanish by reflecting $\lambda\to n-\lambda$, $j\to -j-n$.
Using eqs.~(\ref{a-+}) and (\ref{eq64}), we deduce $\delta Q_n^S = -\delta Q_n^S$, and therefore,
\be \delta Q_n^S =0 \ee
Thus, we arrive at a modified expression
\be\label{eqQn} Q_n^S = \2\sum_{j=-\infty}^\infty\int_0^\infty \frac{d\o}{\sinh\pi\o}\; (i\o +j -nh_-)\; \alpha_{\o+ni\; j-n}^{S\dagger} \alpha_{\o j}^S\ee
which agrees with~(\ref{eqQ}) for $n=-1,0,1$.
We have been cavalier about ordering of modes, since we are discussing classical expressions.
We shall use eq.~(\ref{eqQn}) for all $n\in\mathbb{Z}$.
These conserved charges form a closed algebra.
It is a straightforward exercise to show that
they form a Virasoro algebra,
\be\label{q-algebra}
  [Q_m^S,Q_n^S] = (m-n)Q_{m+n}^S\, ,
\ee
with vanishing classical central charge.

Next, we investigate how the charges act on a scalar field in three-dimensional de Sitter space.  In the bulk (southern diamond) a short calculation shows
that the charges act on the scalar field as
\be   [Q_n^S , \Phi^S] = \zeta_n \Phi^S\ \ ,\ \ n=-1,0,1,
\ee
where $\zeta_n$ are the Killing vectors defined in (\ref{zeta}), as expected.
For all other $n$, we obtain non-local expressions, since the corresponding charges do not generate local symmetries.

Near the boundary $\I^-$,
\be
  [Q_n^S , \Psi^-] = \xi_n
\sum_{j=-\infty}^\infty\int_{-\infty}^\infty \frac{d\o}{\sinh\pi\o}\; \mathcal{F}_{n\o j} B_{\o j}\; \alpha_{\o j}^S \; e^{-i\o t+ij\theta} \;,
\ee
where
\be\xi_n = \2\; e^{-niw} (\p_w - nh_-) \ \ , \ \ \forall n\in\mathbb{Z}\ee
generate conformal transformations on the boundary with $w=\theta + it$ (eq.~(\ref{eqconf})), obeying the algebra
\be [\xi_m , \xi_n ] = (m-n) \xi_{m+n} \ee
and
\be \mathcal{F}_{n\o j} = \left\{ \begin{array}{ccc} 1 & , & \o \ge 0\\ \\
\frac{i\o + j -nh_-}{i\o+j-nh_+}\; e^{2i(\vartheta_{\o j} - \vartheta_{\o +in\; j-n})} & , & \o < 0\end{array}\right. \ee
It is easy to see from eq.~(\ref{eq64}) that
\be \mathcal{F}_{n\o j} = 1 \ \ , \ \ \forall\o \in \mathbb{R} \ \ , \ \ n=-1,0,1\ee
therefore,
\be
  [Q_n^S , \Psi^+] = \xi_n \Psi^+ \ \ , \ \ n=-1,0,1\ee
For other $n$, we may perform a high-frequency expansion.
Using eq.~(\ref{th-+}), we obtain
\be \mathcal{F}_{n\o j} = 1 + o\left( (i\o +j)^{-1} \right) \ , \ee
therefore
\be
  [Q_n^S , \Psi^+] = \xi_n \Psi^+ + \dots \ , \ee
where the corrections vanish at short distances.

\section{Quantization}\label{sect4}

The full quantum theory includes interactions between the scalar field (in general, matter fields) and the gravitational field.
In the weak coupling limit (in which we are working), interactions may be ignored.
It is often stated that matter fields do not contribute to the central charge,
the latter being given by~(\ref{eq0}).
However, the matter fields renormalize Newton's constant, hence also alter the
central charge of pure gravity through (infinite) renormalization.
For example, the component $T_{zz}$ of the scalar field stress-energy tensor
in de Sitter space with metric~(\ref{eqpl}) is modified to
\be
  T_{zz} = (\p_z\Phi)^2 +G \Theta_{zz} \mathcal{L} \ \ , \ \
  \mathcal{L} = \2 g^{\mu\nu} \p_\mu\Phi \p_\nu\Phi - \2 m^2 \Phi^2
\ee
when the metric is perturbed as in~(\ref{g}), and similarly for $T_{\bar z\bar z}$.
The vacuum expectation value of the modification is proportional to $\langle\mathcal{L}\rangle$ and contributes an infinite renormalization to the gravitational field $\Omega$.
Thus, in the quantum theory, the stress-energy tensor of the matter fields cannot be separated in a meaningful way from the stress-energy tensor of the gravitational field.
Consequently, the Noether charges $Q_n$ ($n=-1,0,1$) cannot be independently defined. The physical quantities are
\be\label{eqQph} \mathcal{Q}_n = Q_n + L_n = Q_n^S + Q_n^N + L_n\ee
where $L_n$ is given in bulk form~(\ref{eqLbu}), or by its equivalent boundary
expression~(\ref{eqLb}).
When extended to all $n\in\mathbb{Z}$, the above conserved charges form a Virasoro algebra of central charge given by~(\ref{eq0c}).

On the other hand, because of symmetry, the infinities arising in $Q_n^{S,N}$
in the southern and northern diamonds, respectively, match each other.
Consequently, the charges
\be\label{eqR} \mathcal{R}_n = Q_n^S - Q_n^N\ee
are well-defined and form a Virasoro algebra of vanishing central charge,
\be [ \mathcal{R}_m , \mathcal{R}_n] = (m-n) \mathcal{R}_{m+n} \ee
in the semiclassical approximation we are working in.

We can build the Hilbert space by acting with creation operators on the product of the three vacuum states corresponding to the southern and northern diamonds ($|0\rangle_{S,N}$) and the Liouville field $\Omega$ ($|0\rangle_\Omega$), defined
respectively by
\be \alpha_{\o j}^S |0\rangle_S = 0 \ \ , \ \ \alpha_{-\o\; j}^N |0\rangle_N = 0 \ \ , \ \ \o > 0\ee
Let us define the gravitational vacuum state $|0\rangle_\Omega$ as the $SL(2,\mathbb{C})$ vacuum state satisfying
\be\label{eqgrv} L_n |0\rangle_\Omega = \bar L_n |0\rangle_\Omega = 0 \ \ , \ \ n\ge -1\ee
On the matter side, the $SL(2,\mathbb{C})$ invariant vacuum state is the Euclidean vacuum state defined by
\be
  |E\rangle = C_E\ \exp\left\{ \sum_{j=-\infty}^\infty \int_0^\infty \frac{d\o}{e^{2\pi\o} -1}\; e^{-2i\vartheta_{\o j}}\; \alpha_{-\o \; -j}^S \alpha^N_{\o j}\right\}\;|0\rangle_S\otimes |0\rangle_N
\ee
where we have included an (infinite) normalization constant, $C_E$.
Both $\alpha_{-\o j}^S$ and $\alpha^N_{\o j}$ are creation operators
for $\o >0$,
acting on the vacuum states $|0\rangle_{S,N}$, respectively.
This state is annihilated by the appropriately normalized annihilation operators
\be\label{eqab} \beta_{\o j} = \frac{1}{\sqrt{2\sinh\pi\o}} (e^{\pi\o/2} \alpha_{\o j}^S - e^{-\pi\o/2} \alpha_{\o j}^N)
\ \ , \ \ \o\in\mathbb{R}\ee
It should be emphasized that these are annihilation operators, for both positive and negative frequencies.
They satisfy the commutation relations ({\em cf.}~eq.~(\ref{eqaad}))
\be [\beta_{\o j} , \beta_{\o' j'}^\dagger ] = \sinh\pi\o\; \delta(\o-\o') \delta_{jj'}\ee
We may express the `bulk' $\alpha$-modes in terms of the Euclidean $\beta$-modes by inverting~(\ref{eqab}). We obtain
\bes \alpha_{\o j}^S &=& \frac{1}{\sqrt{2\sinh\pi\o}} (e^{\pi\o/2} \beta_{\o j} - ie^{-\pi\o/2} e^{2i\vartheta_{\o j}} \beta_{-\o\; -j}^\dagger)\nonumber\\
\alpha_{\o j}^N &=& -\frac{1}{\sqrt{2\sinh\pi\o}} (e^{-\pi\o/2} \beta_{\o j} - ie^{\pi\o/2} e^{2i\vartheta_{\o j}} \beta_{-\o\; -j}^\dagger)\ees
We may express the charges in terms of $\alpha$-modes,
\bes
  Q_n &=& Q_n^S + Q_n^N \nonumber\\
&=& \2\sum_{j=-\infty}^\infty\int_0^\infty \frac{d\o}{\sinh\pi\o}\; (i\o+j-nh_-)\left\{ \alpha^{S\dagger}_{\o+in\; j-n} \alpha^S_{\o j} -\alpha^N_{\o+in\; j-n} \alpha^{N\dagger}_{\o j}\right\}
\ees
or equivalently in terms of the Euclidean creation and annihilation $\beta$-modes,
\bes Q_n &=& \2\sum_{j=-\infty}^\infty\int_0^\infty \frac{d\o}{\sinh\pi\o}\; \left\{
(i\o+j-nh_-)\; \beta^\dagger_{\o+in\; j-n}\beta_{\o j} \right. \nonumber\\
 & & \left. \ \ \ \ \ \ \ \ \ \ - (i\o+j+nh_+)\; e^{2i(\vartheta_{\o-ni\; j+n} -\vartheta_{\o j})}\; \beta^\dagger_{-\o+in\; -j-n}\beta_{-\o\; -j}\right\}
\ees
after normal ordering.
In particular, the generators of the $SL(2,\mathbb{C})$ subalgebra take on a simple form,
\be Q_n = \2\sum_{j=-\infty}^\infty\int_{-\infty}^\infty \frac{d\o}{\sinh\pi\o}\; (i\o+j -nh_-)\; \beta^\dagger_{\o+in\; j-n}\beta_{\o j} \ \ , \ \ n=-1,0,1
\ee
on account of eq.~(\ref{eq64}), which however does not hold for any other $n$.
Evidently, all charges annihilate the Euclidean vacuum,
\be Q_n |E\rangle = 0 \ \ , \ \ \forall n\in\mathbb{Z}\ee
This implies (from eqs.~(\ref{eqQph}) and (\ref{eqgrv}))
\be \mathcal{Q}_n\ |E\rangle\otimes |0\rangle_\Omega = 0 \ \ , \ \ n\ge -1 \ee
confirming that this is a $SL(2,\mathbb{C})$ invariant vacuum state for the entire system (matter $+$ gravity).
The other Virasoro generators $\mathcal{R}_n$ (eq.~(\ref{eqR})) may also be expressed in terms of the $\beta$-modes. They contain terms quadratic in creation operators $\beta^\dagger$ and therefore do not annihilate the Euclidean vacuum.

The general wavefunction on the boundary is a linear combination of $\Psi^{S\pm}$ (eq.~(\ref{eqwfb})) and its northern counterpart, $\Psi^{N\pm}$, and can therefore be expressed in terms of the $\alpha$-modes or, equivalently, the $\beta$-modes.
Therefore, all Virasoro generators are expressible in terms of the modes of
boundary wavefunctions, encoding the symmetries of the boundary CFT.
No reference to the southern and northern diamonds in the bulk de Sitter space is needed.
Observables in the bulk may be obtained in terms of the $\beta$-modes.
It would be desirable to study them in detail and develop a physical intuition in terms of the boundary CFT dynamics.

\section{Conclusion}\label{sect5}
We studied the dS/CFT correspondence in three dimensions for a massive scalar field expanding on the resuls of ref.~\cite{BMS}.
We obtained explicit expressions for the generators of isometries of dS$_3$
and brought them to a form that suggested an extension of the algebra to an infinite-dimensional Virasoro algebra.
We studied the action of the Virasoro generators on the boundary wavefunctions and discussed quantization based on the Euclidean vacuum.
We showed that the Virasoro algebra could be defined entirely in terms of the dynamics of the boundary CFT which provided a dictionary for the dS/CFT
correspondence. Thus, dS observables may be understood in terms of corresponding quantities in the boundary CFT. Further work on the CFT side is needed to develop physical intuition for and better understand the dS/CFT correspondence.

 \end{document}